\documentclass[english,preprints,article,accept,moreauthors,pdftex]{Definitions/mdpi}
\usepackage[T1]{fontenc}
\usepackage{textcomp}
\usepackage{amsmath}
\usepackage{amsthm}
\usepackage{graphicx}

\makeatletter

\abstract{Cell-based crowd evacuation systems provide adaptive or static exit-choice
indications that favor a coordinated group dynamic, improving evacuation
time and safety. While a great effort has been made to modeling its
control logic by assuming an ideal communication and positioning infrastructure,
the architectural dimension and the influence of pedestrian positioning
uncertainty have been largely overlooked. In our previous research,
a Cell-based crowd evacuation system (CellEVAC) was proposed that
dynamically allocates exit gates to pedestrians in a cell-based pedestrian
positioning infrastructure. This system provides optimal exit-choice
indications through color-based indications and a control logic module
built upon an optimized discrete-choice model. Here, we investigate
how location-aware technologies and wearable devices can be used for
a realistic deployment of CellEVAC. We consider a simulated real evacuation
scenario (Madrid Arena) and propose a system architecture for CellEVAC
that includes: a controller node, a radio-controlled LED wristband
subsystem, and a cell-node network equipped with active Radio Frequency
Identification (RFID) devices. These subsystems coordinate to provide
control, display and positioning capabilities. We quantitatively study
the sensitivity of evacuation time and safety to uncertainty in the
positioning system. Results showed that CellEVAC was operational within
a limited range of positioning uncertainty. Further analyses revealed
that reprogramming the control logic module through a simulation-optimization
process, simulating the positioning system's expected uncertainty
level, improved the CellEVAC performance in scenarios with poor positioning
systems.}

\providecommand{\tabularnewline}{\\}

\firstpage{1}  
\pubvolume{xx} 
\issuenum{1} 
\articlenumber{5} 
\pubyear{2020} 
\copyrightyear{2020} 
\history{Received: date; Accepted: date; Published: date} 
\pdfoutput=1

\Title{LED wristbands for Cell-based Crowd Evacuation: an Adaptive Exit-choice Guidance System Architecture} 

\Author{Miguel A. Lopez-Carmona $^{*}$\orcidA{} and Alvaro Paricio $^{} $\orcidB{} }  

\AuthorNames{Miguel A. Lopez-Carmona and Alvaro Paricio} 

\address{%
\quad Departamento de Automatica, Escuela Politecnica Superior, Universidad de Alcala, Madrid, Spain; miguelangel.lopez@uah.es (M.A.L.C.); alvaro.paricio@uah.es (A.P.)}

\corres{Correspondence: miguelangel.lopez@uah.es; Tel.: +34-91-885-66-73}  

\makeatother

\usepackage{babel}
\begin{document}
Crowd evacuation; LED wristbands; behavioral optimization; exit-choice
decisions; simulation-optimization modeling; cell-based evacuation

\section{Introduction}

Uncoordinated crowd behaviors are known as being responsible for pedestrians'
deaths and injuries in crowd evacuations. An efficient evacuation
plan is crucial to direct and coordinate evacuees in a safe manner.
This coordination can be achieved by deploying guidance systems that
provide information for each user on the paths, the exit gates, or
the evacuation start time \cite{Abdelghany20141105}.

Many algorithms have been devised for the development of evacuation
guidance systems \cite{biSurveyAlgorithmsSystems2019}. For example,
network flow-based approaches model evacuation planning as a minimum
cost network flow problem \cite{10.2307/j.ctt183q0b4}. The main downside
of network flow-based models is that evacuees must follow the paths
accurately and fulfill an exact schedule. Various approaches have
been suggested to solve this problem using geometric graphs \cite{liLocalizedDelaunayTriangulation2003}.
For example, in \cite{chenDistributedAreaBasedGuiding2008} a wireless
sensor network is partitioned into triangular areas based on the average
detected temperature, and safe egress paths are calculated. Following
this idea, queuing models \cite{newell2013applications} build a queuing
network to estimate evacuation and congestion delays. Various approaches
dynamically develop navigation paths by applying artificial potential
fields to the exits and hazards \cite{koditschekRobotPlanningControl1989,hillSystemArchitectureDirections2000,liDistributedAlgorithmsGuiding2003}.
This technique suffers from several problems, among which is the convergence
time for network stabilization, and its search efficiency in scenarios
with several exit gates. 

There is an extensive research on biologically-inspired algorithms
to search for optimal routes or recommend exits. For instance, in
\cite{liMultiobjectiveEvacuationRoute2010} a multiobjective evacuation
route assignment model based on genetic algorithm \cite{hollandAdaptationNaturalArtificial1992a,gelenbeGeneticAlgorithmsRoute2006}
is proposed. In \cite{Yadegari2010ABO} bee colony optimization is
used to displace evacuees to safe areas. The downside of this work
is the relatively high communication overhead. A wearable device named
LifeBelt is proposed in \cite{Ferscha201033} that recommends exits
to individuals based on the sensed environment. In \cite{wongOptimizedEvacuationRoute2017}
a shortest path algorithm computes pedestrian routes by iteratively
partitioning graph edges at critical division points. Routes are iteratively
refined offline until an optimal state is achieved. This approach
assumes that a crowd distribution is known in advance, and does not
adapt to changes during evacuation.

Since many of existing emergency response systems are built on top
of Wireless Sensor Networks (WSN), routing protocols for packet transmission
have been adapted to develop guidance systems in emergency scenarios.
In \cite{filippoupolitisEmergencySimulationDecision2010} an emergency
support system built on top of WSN is presented, which is inspired
by the cognitive packet network routing \cite{nowakCognitivePacketNetworks2019}
in the IoT domain. Since communications are essential in an evacuation
process, opportunistic communications have also been used in the design
of emergency support systems \cite{gorbilOpportunisticCommunicationsEmergency2011}.

It is well known that the performance of crowd evacuation processes
during emergencies can be strongly affected by exit-choice decision
making at the individual level \cite{kinatederExitChoiceEmergency2018,chenElementaryStudentsEvacuation2018,haghaniSimulatingDynamicsAdaptive2019}.
Thus, there are research efforts in the area of real-time guidance
for crowd evacuations that have focused on studying mechanisms for
providing pedestrians with optimal exit-choice information. A promising
line of research in this area is that of cell-based evacuation systems
\cite{Abdelghany20141105,Zhong2016127,lopez-carmonaCellEVACAdaptiveGuidance2020}.
These systems rely on a cell-based pedestrian positioning infrastructure
such that pedestrians in a cell are assumed to receive the same exit
gate instructions. In \cite{Abdelghany20141105}, a simulation-optimization
framework integrates a genetic algorithm and a microscopic pedestrian
simulation-assignment model. Evacuees are assumed to receive exit-choice
indications that may include the optimal start time of evacuation.
Similarly, in \cite{Zhong20141125} the idea is to use a gene expression
programming to find a heuristic rule. This rule is used to indicate
people in the same sub-region to move towards the same exit. The main
drawback of these approaches is that they do not consider the dynamic
environment features. 

Since the dynamics of the environment change over time in unpredictable
ways, adaptive strategies are recognized as more adequate solutions
\cite{haghaniSimulatingDynamicsAdaptive2019}. There exist adaptive
approaches of cell-based evacuation systems, in which the system's
control logic module updates the cells' exit-choice indications in
real-time depending on the existing environmental conditions. In \cite{Zhong2016127},
they propose a heuristic rule that considers the distance and width
of exit doors as fixed input parameters and density around a given
cell as a dynamic parameter. The crowd evacuation planning problem
is converted to finding the optimal heuristic rule that minimizes
the total evacuation time. To solve this problem, the authors adopt
the Cartesian Genetic Programming (CGP) \cite{Miller2011}. We developed
in \cite{lopez-carmonaCellEVACAdaptiveGuidance2020} an adaptive cell-based
crowd evacuation system (CellEVAC) based on behavioral optimization
that searches for the optimal evacuation plan through meta-heuristic
optimization methodology. As in \cite{Zhong2016127}, we obtain adaptive
evacuation plans capable of responding to changing environmental conditions.
However, our control logic model is easier to configure and optimize,
with a more straightforward logic formulation and interpretation,
exhibiting a more natural pedestrian behavior.

All these approaches outlined above have mainly focused on the design
of algorithms to provide optimal exit-choice indications, assuming
that there exist ideal communication and pedestrian positioning infrastructures.
However, for a real and effective implementation of this type of system,
it is necessary to propose concrete hardware architectures whose deployment
is technologically feasible at a reasonable cost. Also, given an architectural
proposal, it will be essential to evaluate its influence on the performance
of the evacuation processes, and if appropriate, propose corrective
actions for its improvement.

In this work, we are particularly interested in proposing an adaptive
cell-based evacuation system architecture using existing communication
and positioning technologies, paying attention to usability, which
is essential in emergency evacuations where the information of routing
to exit gates should be easily interpretable. Another central question
of this study concerns quantifying the influence of pedestrian positioning
uncertainty in evacuation time and safety. We would also like to quantify
the importance of reprogramming the control logic module under uncertainty
conditions by using simulation-optimization techniques. Given a control
logic optimized assuming an error-free positioning system, what would
be the quantitative benefit of re-optimizing the control logic if
we assume a level of uncertainty in pedestrian positioning.

With the purposes mentioned above, this paper proposes a system architecture
for our adaptive cell-based evacuation system CellEVAC \cite{lopez-carmonaCellEVACAdaptiveGuidance2020}.
The proposed system architecture consists of three subsystems: (i)
monitoring and control logic subsystem, (ii) active RFID cell-node
network, and (iii) radio-controlled LED wristband subsystem. 

The RFID cell-node network and radio-controlled LED wristband subsystems
coordinate to provide exit gate indication display and positioning
capabilities to pedestrians. The monitoring and control logic subsystem
monitors the environmental conditions and accordingly allocates exit
gate colors to cell-nodes in real-time. Thus, the LED wristbands show
the color corresponding to the cell in which each pedestrian is located,
indicating the recommended exit gate. 

In \cite{lopez-carmonaCellEVACAdaptiveGuidance2020}, we assumed an
error-free positioning infrastructure, where pedestrians were supposed
to be equipped with a generic device (smart-phone or wearable device)
with ideal location-aware and color display capabilities. In this
paper, the proposed positioning system's RFID communication channels
are modeled using a log-normal propagation model. To define different
uncertainty levels in pedestrian positioning and study its influence
in evacuation time and safety, we modulate the Gaussian distribution
that models the random variations in the propagation model. Finally,
we apply the simulation-optimization methodology to obtain the control
logic subsystem's optimal configuration under different uncertainty
levels. This approach gives us information about the importance of
reprogramming the control logic if we know in advance the positioning
uncertainty level.

To perform simulation and simulation-optimization experiments, we
have used the simulation-optimization modeling framework that we developed
in \cite{lopez-carmonaCellEVACAdaptiveGuidance2020}. This framework
integrates a microscopic pedestrian simulation based on the classical
Social Force Model (SFM) \cite{helbingSocialForceModel1995}. The
simulation-optimization process adopts a Tabu-Search algorithm (TS)
\cite{fredgloverTabuSearch1997}, which iteratively searches for the
near-optimal evacuation plan. At the same time, the microscopic crowd
simulation guides the search by evaluating the evacuation time and
safety of the solutions generated by the TS algorithm. 

The rest of the paper is organized as follows. Section \ref{sec:Simulation-model}
presents our proposal of system architecture for CellEVAC, the microscopic
simulation-optimization framework used to perform the experimental
evaluation, and the mechanism to model positioning uncertainty. Section
\ref{sec:Experiments} presents the experimental evaluation and results.
The last section provides concluding comments and possible research
extensions.

\section{\label{sec:Simulation-model}Methods}

\subsection{Evacuation Scenario}

\begin{figure}
\begin{centering}
\includegraphics[width=0.9\columnwidth]{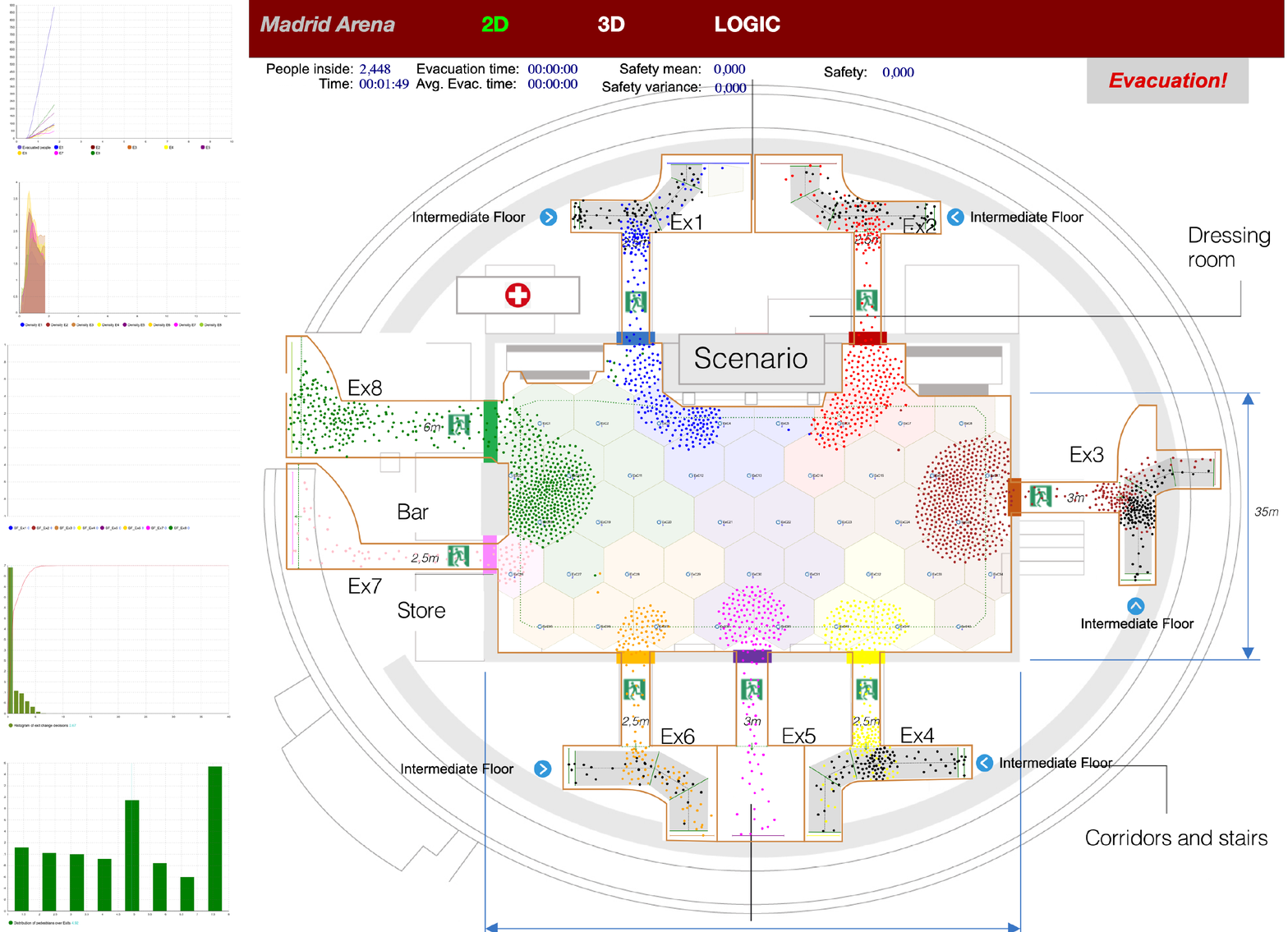}
\par\end{centering}
\caption{\label{fig:Madrid-Arena-layout.}Madrid Arena layout (ground floor).}
\end{figure}

Our evacuation scenario was \emph{Madrid Arena}, an indoor arena located
in Madrid that hosts sports events, commercial, cultural and leisure
activities. It has three floors (access, intermediate, and ground)
and 30,000 m\texttwosuperior , with a capacity of 10,248 spectators.
We studied the evacuation of the ground floor, which has a maximum
capacity of $3,400$ spectators with its retractable bleachers removed.
Figure \ref{fig:Madrid-Arena-layout.} 
 shows the ground floor layout, with $1,925m^{2}$ and eight exit
gates (\emph{Ex1} to \emph{Ex8}) with widths in the range $2.5m$
and $6m$. Pedestrian flows from intermediate floors were simulated
by injecting pedestrians at exits 1, 2, 3, 4, and 6 at the entry points
highlighted with a blue dot. 

As in \cite{lopez-carmonaCellEVACAdaptiveGuidance2020}, we divided
the ground floor into 42 regular hexagonal cells of $9m^{2}$ and
$6m$ width, whose dimensions were chosen to provide a balance between
control, wireless coverage, and computational efficiency. 

\subsection{System Architecture for CellEVAC}

\begin{figure}
\begin{centering}
\includegraphics[width=0.5\paperwidth]{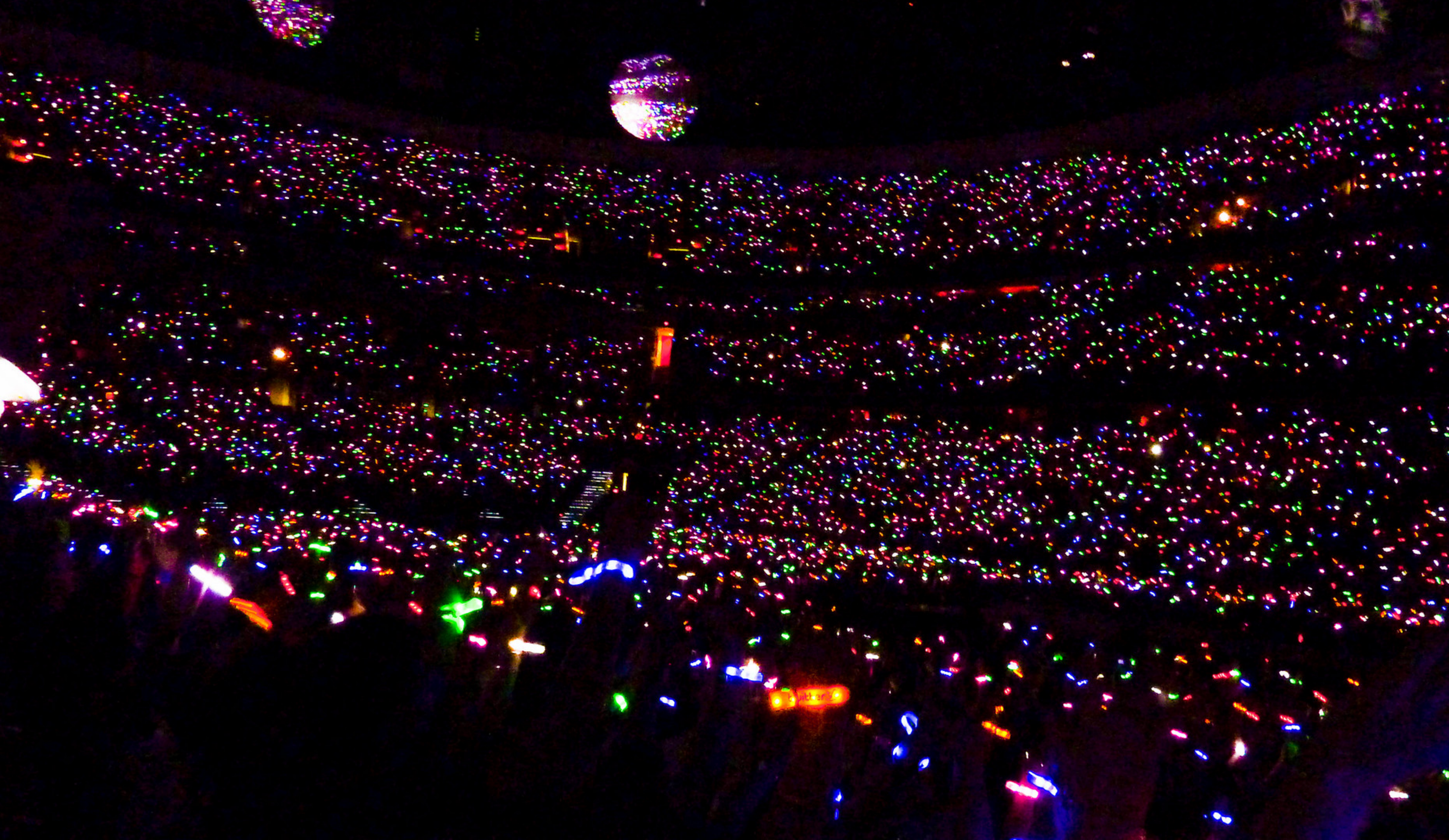}
\par\end{centering}
\caption{\label{fig:Concert-of-Coldplay}Concert of Coldplay at Verizon Center.
Spectators wear LED wristbands manufactured by Xylobands (http://xylobands.com).
Author: Matthew Straubmuller; license: https://creativecommons.org/licenses/by/2.0/;
source: https://www.flickr.com/photos/imatty35/7550673548.}
\end{figure}

We considered using radio-controlled LED wristbands that display colors
recommending an exit gate. These LED wristbands are widely used at
a range of events, from live acts at arenas to product launches, sports
matches, parties, and corporate events from 1-150,000 people. The
wristbands work by creating multiple flash patterns with RGB LEDs
that use the full-color spectrum and can be programmed to create visual
effects (Figure \ref{fig:Concert-of-Coldplay}). Xylobands (http://xyloband.com)
or CrowdLED (https://crowdled.net) are two examples of companies offering
these kinds of products. Usually, radio control has a range of hundreds
of meters, and the wristbands have a battery life of approximately
20 hours. Two extended features that can be found are the inclusion
of RFID tags for registration purposes and zonal control to activate
wristbands in separate groups. 

Our idea was to extend the functionality of these devices, which is
oriented towards creating visual effects, using them in case of emergency
to guide people to color-illuminated exit gates. The displayed color
in the wristband indicates the evacuee the corresponding exit gate.
Besides, a synergic effect of using LED wristband lightning is that
it may ease image processing for pedestrian flow estimation, which
is used in our system to build the control logic.

\begin{figure}
\begin{centering}
\includegraphics[width=0.45\linewidth]{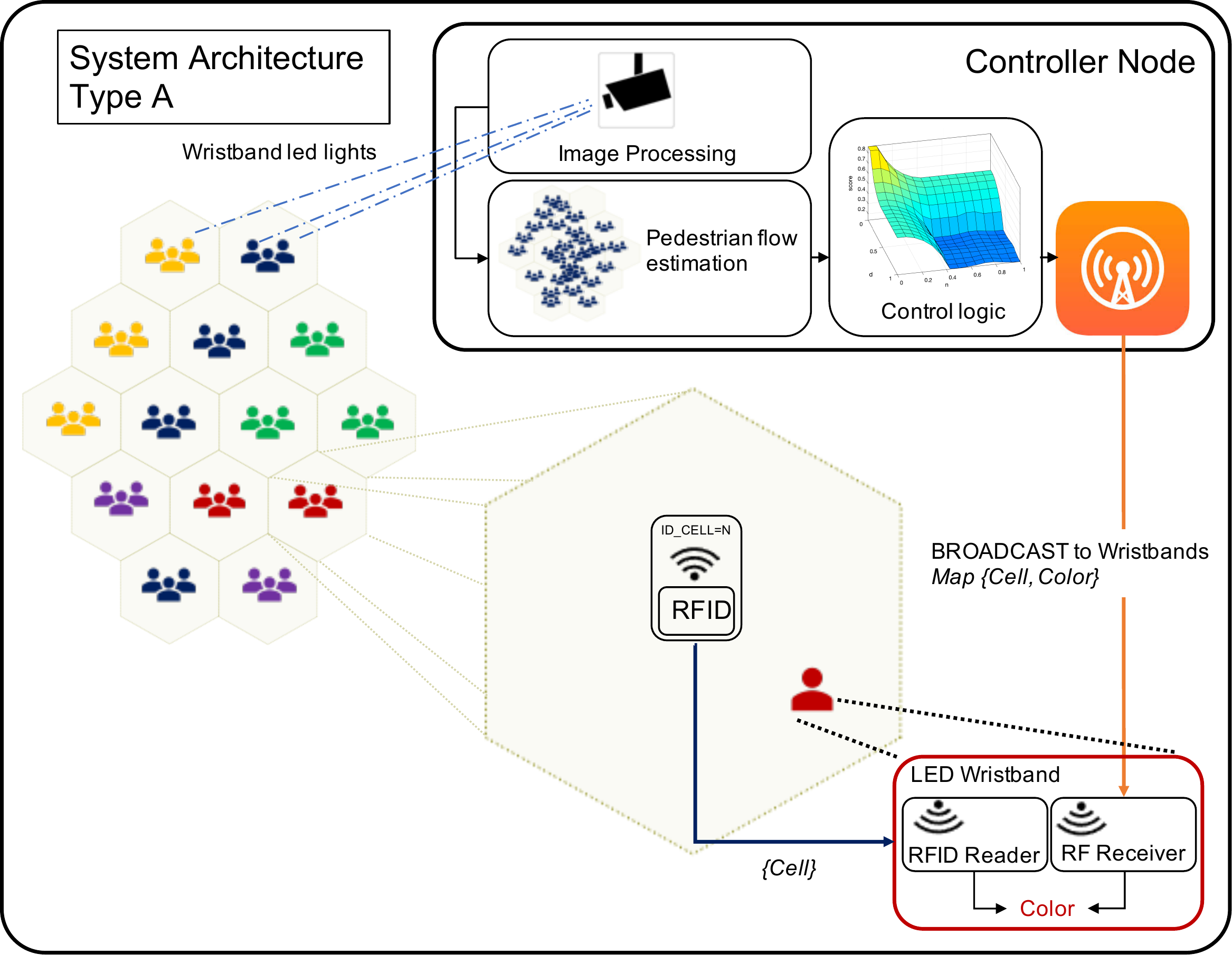}
\par\end{centering}
\begin{centering}
\includegraphics[width=0.45\linewidth]{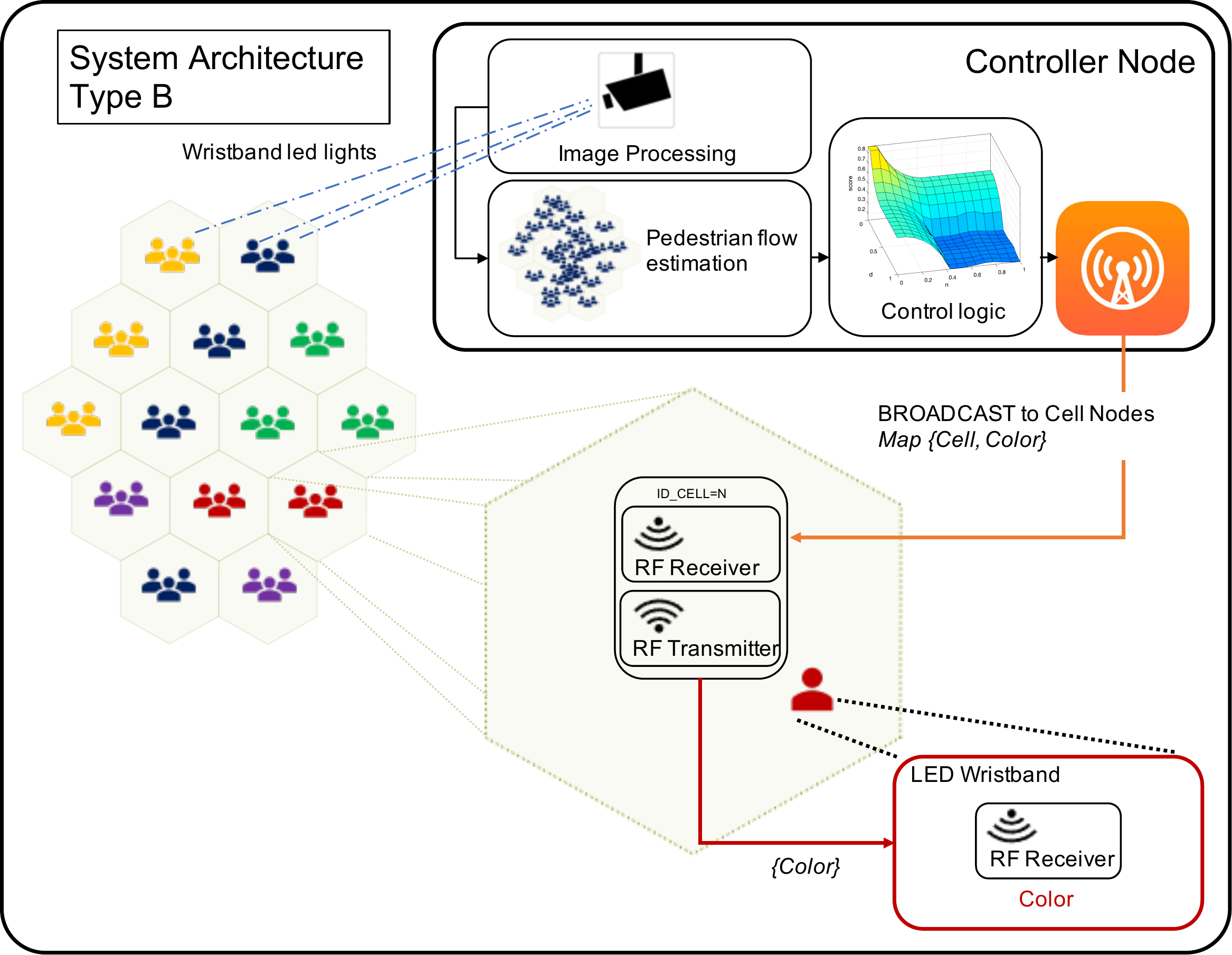}
\par\end{centering}
\caption{\label{fig:CellEVAC-System-Architecture}CellEVAC System Architecture:
Types A and B.}
\end{figure}

As described in the introduction section, the proposed system architecture
consists of three subsystems: 
\begin{enumerate}
\item Monitoring and control logic subsystem (Controller Node), which monitors
pedestrian flows using image processing and generates exit-choice
indications in the form of color allocation to cells.
\item Active RFID cell-node network whose purpose is to provide positioning
information to pedestrians' LED wristbands.
\item Radio-controlled LED wristband subsystem, which includes the LED wristbands
with color display and radio-frequency (RF) communication capabilities. 
\end{enumerate}
Figure \ref{fig:CellEVAC-System-Architecture} shows two possible
implementations of the system architecture (Types A and B) for deploying
CellEVAC using existing off-the-shelf technologies. Both alternatives
install a controller node with three functional blocks: pedestrian
flow estimation based on image processing, control logic based on
behavioral optimization \cite{lopez-carmonaCellEVACAdaptiveGuidance2020},
and RF transmitter. 

In the controller node, the pedestrian flow estimation block performs
image processing to detect LED wristbands lightning and estimate pedestrian
density at each cell. In this work we assumed a flow estimation block
that is based on commercially available pedestrian counting technology
\cite{lesaniDevelopmentEvaluationRealtime2020,kurkcuEstimatingPedestrianDensities2017,akhterIoTEnabledIntelligent2019,ilyasConvolutionalNeuralNetworkBasedImage2020}.
Obtained pedestrian densities feed the control logic block that computes
the optimal allocation of colors to cell-nodes (see Section \ref{subsec:Pedestrian-behavior-modeling}).
The RF transmitter broadcasts messages periodically containing the
42 tuples $\{Cell,Color\}$ that assigns a color to each cell. This
process repeats every five seconds.

In the Type A architecture, each cell-node is equipped with an active
RFID tag \cite{chenSelfpoweredSmartActive2019} that periodically
broadcasts its ID (active RFID beacon \cite{correaReviewPedestrianIndoor2017}).
On the other hand, the wristbands embed an RFID reader that reads
the IDs from the surrounding cells. The wristband selects the ID of
the message with the highest Received Signal Strength Indicator (RSSI)
to estimate the right pedestrian location \cite{alvarezlopezReceivedSignalStrength2017}.
The other element in the wristband is the RF Receiver, which periodically
evaluates the broadcast messages with the tuples $\{Cell,Color\}$
from the controller node. By matching the wristband location (selected
cell ID) and cell-color tuples, the wristband lights up with the exit
gate color assigned to the cell.

In the Type B architecture, the RF Receiver in the cell-node receives
the broadcast messages from the controller node with the assigned
color. Then, the cell-node broadcasts the corresponding color message,
which is read by the wristbands. As in the Type A architecture, several
broadcast messages from different cell-nodes can be received within
a window time. So, the same signal strength selection mechanism is
used by the wristbands to select the right color.

The most critical part of this architecture is in the positioning
functionality. Both RFID and RF communication channels between the
cell-node network and wristbands have to cope with a complex signal
propagation environment. However, the system does not need to obtain
exact position coordinates but select the right cell in which the
pedestrians are located. It means that a significant lower location
resolution is needed and that there is no need to implement triangulation
mechanisms based on RSSI \cite{brchanRealtimeRFIDLocalization2012}.
Another problem to solve is co-channel interference, which may be
managed using existing radio resource management schemes \cite{assarianEfficientResourceAllocation2019}.
Besides, the RF transmission channel in the controller node is a one-to-many
communication channel that has been used to control commercially available
LED wristbands in large events for more than a decade, and do not
pose a particular challenge. 

\subsection{\label{subsec:Pedestrian-behavior-modeling}Control Logic of CellEVAC}

The control logic of CellEVAC is based on a behavioral optimization
approach proposed in our previous work in \cite{lopez-carmonaCellEVACAdaptiveGuidance2020}.
Here we recall the main concepts that build its operation.

Pedestrians' exit-choice decision modeling based on discrete choice
theory \cite{duivesExitChoiceDecisions2012} inspired the control
logic developed for CellEVAC. Thus, we modeled exit gate color allocation
to cell-nodes using the simplest and most popular practical discrete
choice model, the Multinomial Logit Model (MLM) \cite{duivesExitChoiceDecisions2012,ortuzarModellingTransport2011}.
In the MLM control logic, the controller node allocates exit gates
(colors) to cells using a probabilistic model, in which the allocation
probabilities of exit gate $j$ to cell-node $c$ are given by:

\begin{equation}
P_{cj}=\frac{\exp(V_{cj})}{\sum_{E_{i}\in E(c)}\exp(V_{ci})}\label{eq:Probabilistic-1}
\end{equation}
where $\mathbf{E}=\{E_{i=1...42}\}$ is the set of exit gates, and
$V_{cj}$ is the systematic utility function for cell $c$ and exit
gate $j$, which is given by: 

\begin{alignat}{1}
V_{\mathrm{\mathit{cj}}} & =\beta_{D}\times\frac{DISTANCE_{cj}}{max(DISTANCE)}+\beta_{W}\times\frac{WIDTH_{j}}{max(WIDTH)}\label{eq:Utility function}\\
 & +\beta_{G}\times\frac{GROUP_{cj}-GROUP_{min}}{GROUP_{cj}}+\beta_{E}\times\frac{EXCON_{j}}{criticalDensity_{j}}\nonumber \\
 & +\beta_{C}(t)\times NOCHANGING{}_{cj}\nonumber 
\end{alignat}
\\

The first attribute is the distance from cell-node $c$ to exit gate
$j$, while the second attribute represents the width of each exit
gate. Both attributes are normalized in the range of 0-1.

The third attribute is the $GROUP$ ratio, which estimates the congestion
along a path from cell $c$ to an exit gate $j$, relative to the
least congested path. A group ratio value of 0 indicates that the
chosen path is the least congested. When the value of the group ratio
tends towards 1, it means that the emptiest path's imbalance becomes
large. The parameter $\beta_{G}$ is expected to be positive if it
favors pedestrians to follow other pedestrians and is negative otherwise.
Note that with this normalization, we assume that the attribute's
relevance is kept constant throughout the evacuation process.

The fourth attribute $EXCON$ accommodates the congestion at exit
gates. For a given density value, congestion is higher if the pedestrian
flow is low. We reflect this effect through critical density values
obtained from the fundamental diagrams of each exit gate (see \cite{lopez-carmonaCellEVACAdaptiveGuidance2020}).
This $criticalDensity_{j}$ value reflects the density value at which
the exit gate's maximum capacity is reached. Therefore, the $EXCON_{j}$
value representing density at exit gate $j$ is normalized by the
corresponding $criticalDensity_{j}$ value. This normalization converts
$EXCON$ into a unitless attribute around 1. When the value of $EXCON$
is above 1, it means that exit is highly congested. A value close
to 0 would indicate that the exit gate is almost empty. In contrast
to the normalization procedure used for the $GROUP$ attribute, the
distribution of $EXCON$ values exhibits a decreasing evolution as
the number of pedestrians in the evacuation scenario decreases. It
seems reasonable to assume that the relevance of congestion at exits
as a discriminant factor for exit-choice decreases when the overall
number of pedestrians is low, and so $EXCON$ is close to 0 at all
exits. 

The fifth attribute is the $NOCHANGING$ value associated with cell-node
$c$ and exit $j$, which captures how the controller is likely to
revise the previous exit gate allocation (this attribute was named
$PERSONAL$ in \cite{lopez-carmonaCellEVACAdaptiveGuidance2020})
. We treat this attribute as a binary categorical 0-1 value that equals
1 if the current exit gate of cell $c$ is $j$, and is 0 otherwise
($NOCHANGING=0\:\forall k\neq j$ ). Therefore, in a general context,
the parameter $\beta_{C}(t)$ is expected to be positive if the controller
tends to maintain the previous exit gate allocations, and is negative
otherwise. However, we aimed to modulate the tendency to maintain
previous decisions, and so, $\beta_{C}(t)$ is always positive. As
was noted above, we assumed that exit-choice changing evolves as evacuation
progresses, and therefore the parameter that modulates $NOCHANGING$
is time-dependent. By observing the pattern of behavior under various
simulation settings, it was found reasonable that the tendency to
maintain decisions increased linearly depending on the current number
of pedestrians:
\begin{equation}
\beta_{C}(t)=\beta_{C}\times\left(1-\frac{numOfPeds(t)}{numOfPeds(t=0)}\right)\label{eq:Beta-N}
\end{equation}

According to Equation \ref{eq:Beta-N}, the parameter $\beta_{C}(t\rightarrow0)$
tends to $0$ at the beginning of the evacuation, and so, the tendency
to revise decisions is higher. As the number of pedestrians decreases,
the parameter $\beta_{C}(t)$ tends to $\beta_{C}$, and the tendency
is to maintain previous decisions proportionally to the $\beta_{C}$
value.

In the simulation setting used in this work, we used an update cycle
of 5 seconds. We kept this frequency constant and controlled the frequency
of the changes at optimal levels using the parameter $\beta_{C}$.

\subsection{\label{subsec:Modeling-Positioning-Uncertainty}Modeling Positioning
Uncertainty}

Active RFID systems are defined by three parts, a reader (wristband),
antennas, and a tag (cell-node), with their power source. In active
RFID applications, RSSI can be used for determining the location of
a tag, such that each tag's RSSI value is proportional to the distance.
In our system, the cell-node embeds an active beacon tag that sends
out its ID every 3 - 5 seconds. Thus, each tag\textquoteright s RSSI
value is proportional to the distance between the wristband and cell-node.
However, the RSSI value in active RFID applications can be affected
by multipath and signal collision \cite{brchanRealtimeRFIDLocalization2012}. 

In free-space, the relationship of the power transmitted from cell-node
to wristbands, assuming the antennas are isotropic and have no directivity,
is expressed by the free-space path loss equation derived from the
Friis transmission equation: 
\begin{equation}
PL(dB)=20\log_{10}(d)+20\log_{10}(f)-27.55\label{eq:PL}
\end{equation}
where $PL$ is the free-space path loss in $dB$, $f$ is the signal
frequency in MHz, and $d$ is the distance in meters from the cell-node
to the wristband. For converting RSSI values into a distance measurement
in indoor environments with random shadowing effects, one of the most
common approaches taken is to use the log-normal propagation model
\cite{rappaportWirelessCommunicationsPrinciples2002,xuSpatialSignalAttenuation2018}:
\begin{equation}
P_{RX_{dBm}}=RSSI=P_{TX_{dBm}}-PL_{0}-10\eta\log_{10}\frac{d}{d_{0}}+X_{g}\label{eq:LPM}
\end{equation}
where $P_{TX_{dBm}}$ is the transmitted power in $dBm$, $P_{RX_{dBm}}$
is the received power, $PL_{0}$ is the path loss for a reference
distance $d_{0}$ calculated using the free-space path loss equation
(Equation \ref{eq:PL}) or by field measurements, $d\geq d_{0}$ is
an arbitrary distance, $\eta$ is the path loss exponent, and $X_{g}$
is a gaussian random variable with zero mean and variance $\sigma^{2}$
that models the random variation of the RSSI value. The path loss
exponent $\eta$ in indoor environments such as stadiums can reach
values in the range of 4 to 7. 

User preference or environmental considerations usually prescribe
which parameter configuration to use for most applications. In our
simulation scenario, we used a frequency of $2.45GHz$, transmission
power of $10W$, path loss exponent $\eta=5$, and reference distance
$d_{0}=1m$. Thus, RSSI can be expressed as
\begin{align}
RSSI & =-60\log_{10}(d)+X_{g},\,d\geq1m\label{eq:RSSI}
\end{align}

Modifying the variance $\sigma_{g}^{2}$ of the gaussian distribution
$X_{g}$ we may modulate positioning uncertainty. 

The procedure to calculate each pedestrian's location in evacuation
simulations is a two-step process that repeats every five seconds:
\begin{enumerate}
\item Given the set of cell-nodes $\{c_{i=1...42}\}$ obtain the set of
distances $\{d_{i=1...42}\}$ from pedestrian to each cell-node $c_{i}$,
and calculate the corresponding set \{$RSSI_{i}\}$ using Equation
\ref{eq:RSSI}. 
\item If there exists a distance value $d_{i}$ in set $\{d_{i=1...42}\}$
such that $d_{i}<1m$, the pedestrian location is $c_{i}$. Otherwise,
the pedestrian location corresponds to the cell $c_{i}$ with the
maximum $RSSI_{i}$ value.
\end{enumerate}

\subsection{Microscopic Simulation-optimization Framework}

Much of the related work on crowd evacuations rely on detailed simulations.
We opted for a multi-agent microscopic simulation framework based
on a Social Force Model (SFM) \cite{helbingSocialForceModel1995}
due to its flexibility and ease of integration of complex interaction
and behavior models. Our simulation framework integrates the potential
of SFM to mimic physical interactions among evacuees, and of multi-agent
systems to simulate complex behaviors and interactions \cite{lopez-carmonaCooperativeFrameworkMediated2017}.

In this work, the simulation-optimization software framework we developed
in \cite{lopez-carmonaCellEVACAdaptiveGuidance2020} has been extended
with the positioning uncertainty model. The software framework embeds
agent-based simulation and discrete event simulation, integrating
pedestrian behavior modeling, SFM for pedestrian motion, control logic
of exit gate indications, positioning, and optimization features. 

We used the commercially available programming, modeling and simulation
software packages AnyLogic \footnote{https://www.anylogic.com/ Accessed 18 August 2020}
and Matlab \footnote{https://www.mathworks.com/ Accessed 19 June 2020}.
The kernel of the simulation-optimization software framework is AnyLogic,
which integrates three different modeling methods: discrete event
simulation, agent-based simulation, and system dynamics, built on
top of a Java-based software development framework. The evacuation
scenario layout, pedestrian motion, and evacuation measurements run
in AnyLogic, while exit-choice decisions and control logic of exit
gate indications are implemented in Matlab. AnyLogic and Matlab are
interconnected in a master-slave configuration through the interface
with external Java libraries provided by AnyLogic and the Matlab Java
API engine (see details below). 

\begin{figure}
\begin{centering}
\includegraphics[width=1\columnwidth]{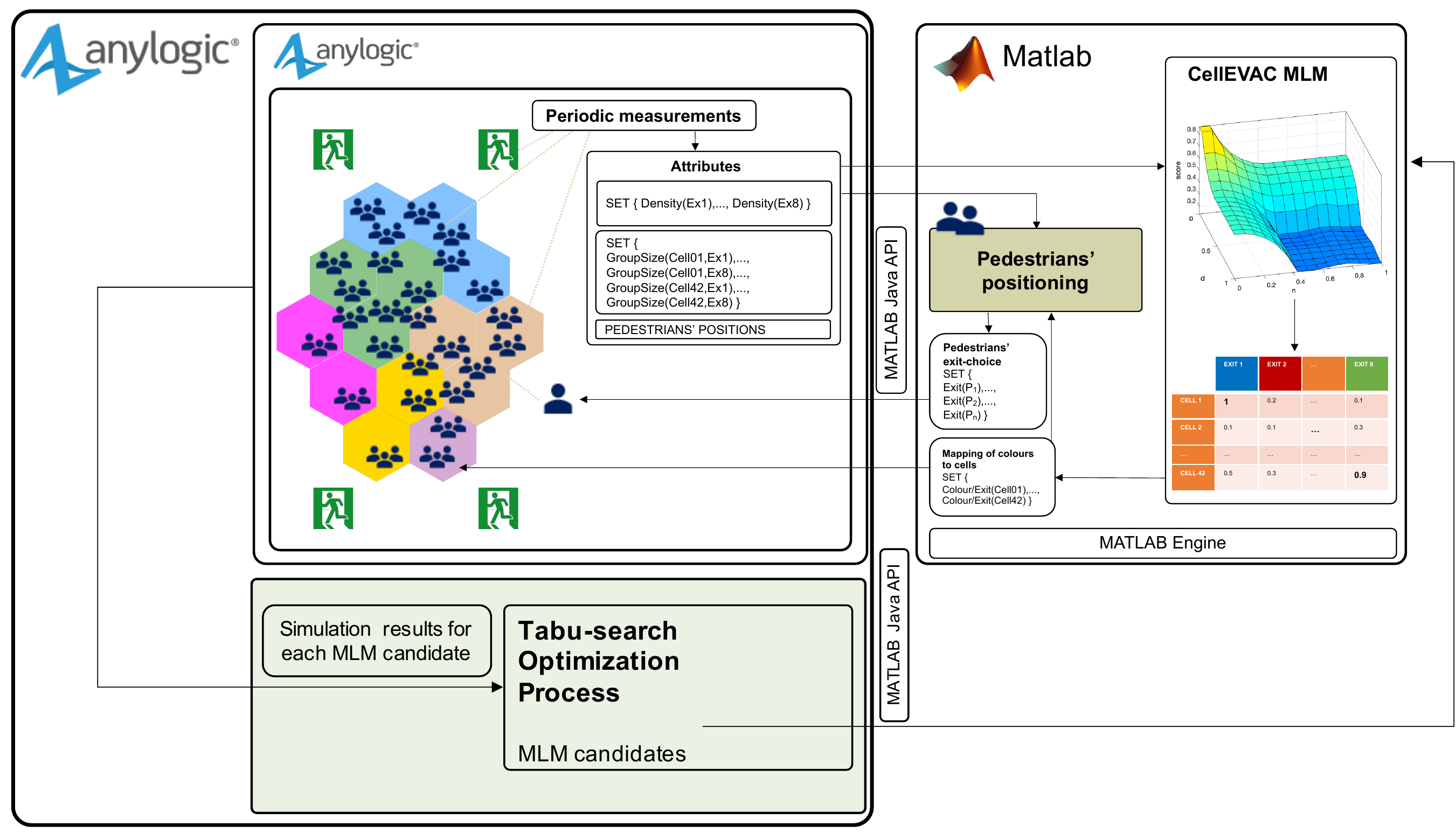}
\par\end{centering}
\caption{\label{fig:SimulationModelCellEVACMLM}Simulation-optimization software
framework of CellEVAC with control logic based on Multinomial Logit
Model (MLM).}
\end{figure}

The CellEVAC simulation model with MLM control logic is shown in Figure
\ref{fig:SimulationModelCellEVACMLM}. The evacuation scenario layout,
visualization features, and all the functionality regarding the SFM
based pedestrian motion were implemented within AnyLogic. 

During a simulation, the first step is to send from AnyLogic to Matlab
the set of parameters that configure the CellEVAC MLM and Pedestrians'
positioning modules, including the set of cell-node center coordinates
and exit gates, and the uncertainty level. Next, the pedestrian positioning
and densities at exit gates and cells are periodically measured and
then transformed into the set of attributes: pedestrian positions,
density at each exit gate, and group of pedestrians along the path
to each exit. The group size of each pair cell-exit gate is calculated
by adding the pedestrians in the cells that are closer to each exit.
All these attributes feed the CellEVAC MLM module in Matlab that implements
the decision logic to map colors (exit gates) to cells. This mapping
is sent back to AnyLogic for visualization purposes, and to the Pedestrians'
positioning module within Matlab to allocate exit gates (colors) to
pedestrians (LED wristbands). The Pedestrians' positioning module
implements the function that locates each pedestrian in a given cell-node
using the positioning uncertainty model. The output of the Pedestrians'
positioning module is the set of pairs pedestrian-exit gate, which
is sent to AnyLogic for simulating pedestrian motion.

To search for optimal configurations of the MLM model, we used a simulation-optimization
process that adopts a Tabu-Search algorithm (TS) \cite{fredgloverTabuSearch1997},
which iteratively searches the solution space. At the same time, the
microscopic crowd simulation guides the search by evaluating the evacuation
time and safety of the solutions generated by the TS algorithm. The
optimization process is built on top of the OptQuest \footnote{https://www.opttek.com/ Accessed 18 August 2020}
optimization engine provided by AnyLogic. Figure \ref{fig:SimulationModelCellEVACMLM}
shows the optimization module on a green background. The parameters
of the CellEVAC MLM model are the ``MLM candidates'' generated by
the TS algorithm. Thus, each candidate is defined by a tentative set
of parameters $\beta$ sent to the MATLAB Engine at each iteration
of the optimization process. The simulations results are sent back
to the optimization module for its evaluation and thus guide the optimization
process.

\section{\label{sec:Experiments}Simulation-optimization Experiments and Results}

The performance measurements in all the experiments were the\emph{
total evacuation time}, \emph{average safety, variance of safety},
and the \emph{average number of pedestrians' exit-choice decision
changes}. The average and variance of safety are based on the safety
values computed at the differents exit gates. Average safety characterizes
the overall safety value, while the variance of safety is used to
estimate the imbalance of safety between the exit gates. The procedure
to calculate exits' safety values is first to obtain their Fundamental
Diagrams (FD) derived through microscopic simulation. The FD represents
the relation between pedestrians' flow and density. Given the FDs,
a procedure is defined to obtain three density thresholds. These density
thresholds and the measurements of density during an evacuation process
are used to calculate the safety values. For a detailed description
of how we modeled pedestrian flows and safety, see \cite{lopez-carmonaCellEVACAdaptiveGuidance2020}.
In this paper, we used the same thresholds and parameters defined
in \cite{lopez-carmonaCellEVACAdaptiveGuidance2020} to measure the
safety values.

We conducted two types of experiments: (i) sensitivity analysis to
positioning uncertainty, and (ii) simulation-optimization. In all
the simulation setups, the evacuee population consisted of 3400 pedestrians
on the ground floor, who had a preferred evacuation speed obtained
from a uniform distribution between 1.24 and 1.48 $m/s$. To speed
up the simulation-optimization experiments, we imposed a deadline
of $25$minutes to each evacuation simulation iteration, after which
the simulation iteration was aborted. 

Two different evacuation scenarios were considered depending on the
experiment: evacuations without external flows (NEF) in which no pedestrians
were coming from the upper floors, and evacuations with external flows
(EF) (i.e., with pedestrians coming from the upper floors) to simulate
more complex pedestrian flow interactions. In EF scenarios, three
exit gates were chosen at random at each simulation iteration. Two
of these exit gates received an incoming pedestrian flow rate of $120peds/m$,
while the third exit gate was blocked. 

In the sensitivity analysis experiments, each experiment ran the evacuation
simulation model multiple times varying the positioning uncertainty
level (variance of the Gaussian distribution $X_{g}$ ), showing how
the simulation output (i.e., the performance measurements) depended
on it. Due to the stochastic nature of the evacuation processes, we
used a replication algorithm to obtain representative results for
a given parameter setting and a specific simulation output. This algorithm
defines a minimum and a maximum number of experimental runs per parameter
setting (replications of a simulation), a confidence level for the
sample mean of replications (simulation output average), and an error
percent. The minimum number guarantees the minimum number of replications,
while the confidence level and error percent determine if more replications
are needed. Simulations for a given parameter configuration stops
when the maximum number of replications has been run or when the confidence
level is within the given percentage of the mean of the replications
to date. In our experimental setup, evacuation time was used as an
output parameter to control the number of replications between 10
and 50. The confidence level was fixed to $95\%$, and the error percent
to $0.5$.

In the simulation-optimization experiments, we used the Tabu-search
optimization algorithm \cite{fredgloverTabuSearch1997}. The goal
was to find the optimal combination of parameters of the MLM model
that resulted in the best possible solution. We considered two different
optimization scenarios characterized by the fitness function (objective
function) used and the existence of external pedestrian flows. 
\begin{itemize}
\item \textbf{NEF}: Optimize Time and Safety ($min(evacTime-Sf)$) without
External Flows. The goal is to optimize evacuation time and average
safety, and the evacuation scenario does not include external pedestrian
flows.
\item \textbf{EF}: Optimize Time and Safety ($min(evacTime-Sf)$) with External
Flows. The goal is to optimize evacuation time and average safety,
and the evacuation scenario includes external pedestrian flows.
\end{itemize}
As in the sensitivity analysis experiments, the optimization algorithm
applies a replication algorithm. However, while in the sensitivity
analysis, the number of replicas was limited by the evacuation time
value, in simulation-optimizations, the stop condition was controlled
by the fitness function (objective function).

\subsection{\label{subsec:Sensitivity-Analysis-of}Sensitivity Analysis of Positioning
Uncertainty}

In the sensitivity analysis of evacuation performance to positioning
uncertainty, the standard deviation $\sigma_{g}$ in $X_{g}$(Equation
\ref{eq:RSSI}) was varied from $0$dB to $40$dB at discrete steps
in two different evacuation scenarios, with and without external pedestrian
flows. To evaluate up to which uncertainty level is beneficial CellEVAC
in comparison with not using a guidance system, we also included the
case in which pedestrians did not use the CellEVAC system (coded as
$\sigma_{g}=N$ in the result box-plots). The optimal parameter configuration
of the MNL model found in \cite{lopez-carmonaCellEVACAdaptiveGuidance2020}
was used to implement the CellEVAC decision logic (Table \ref{tab:Optimal-configuration-of-CellEVAC})..
For the experiments in which pedestrians did not use CellEVAC, the
decision logic was implemented at a pedestrian level using the configuration
of parameters of the MLM model defined in \cite{lopez-carmonaCellEVACAdaptiveGuidance2020}
that simulates standard pedestrian behavior.

\begin{table}
\caption{\label{tab:Optimal-configuration-of-CellEVAC}Optimal parameter configuration
for the CellEVAC MLM decision logic model, and parameter configuration
of the MLM pedestrians' standard behavior model.}

\centering{}%
\begin{tabular}{cccccc}
\hline 
 & {\small{}$\beta_{D}$} & {\small{}$\beta_{G}$} & {\small{}$\beta_{E}$} & {\small{}$\beta_{W}$} & {\small{}$\beta_{C}$}\tabularnewline
\hline 
{\small{}Optimal CellEVAC for 0dB} & {\small{}-17.723} & {\small{}-2.181} & {\small{}-1.671} & {\small{}1.064} & {\small{}2.594}\tabularnewline
{\small{}Standard (without CellEVAC)} & {\small{}-28} & {\small{}0.6} & {\small{}-0.5} & {\small{}0.6} & {\small{}0}\tabularnewline
\hline 
\end{tabular}
\end{table}

\begin{figure}
\begin{centering}
\includegraphics[width=0.8\linewidth]{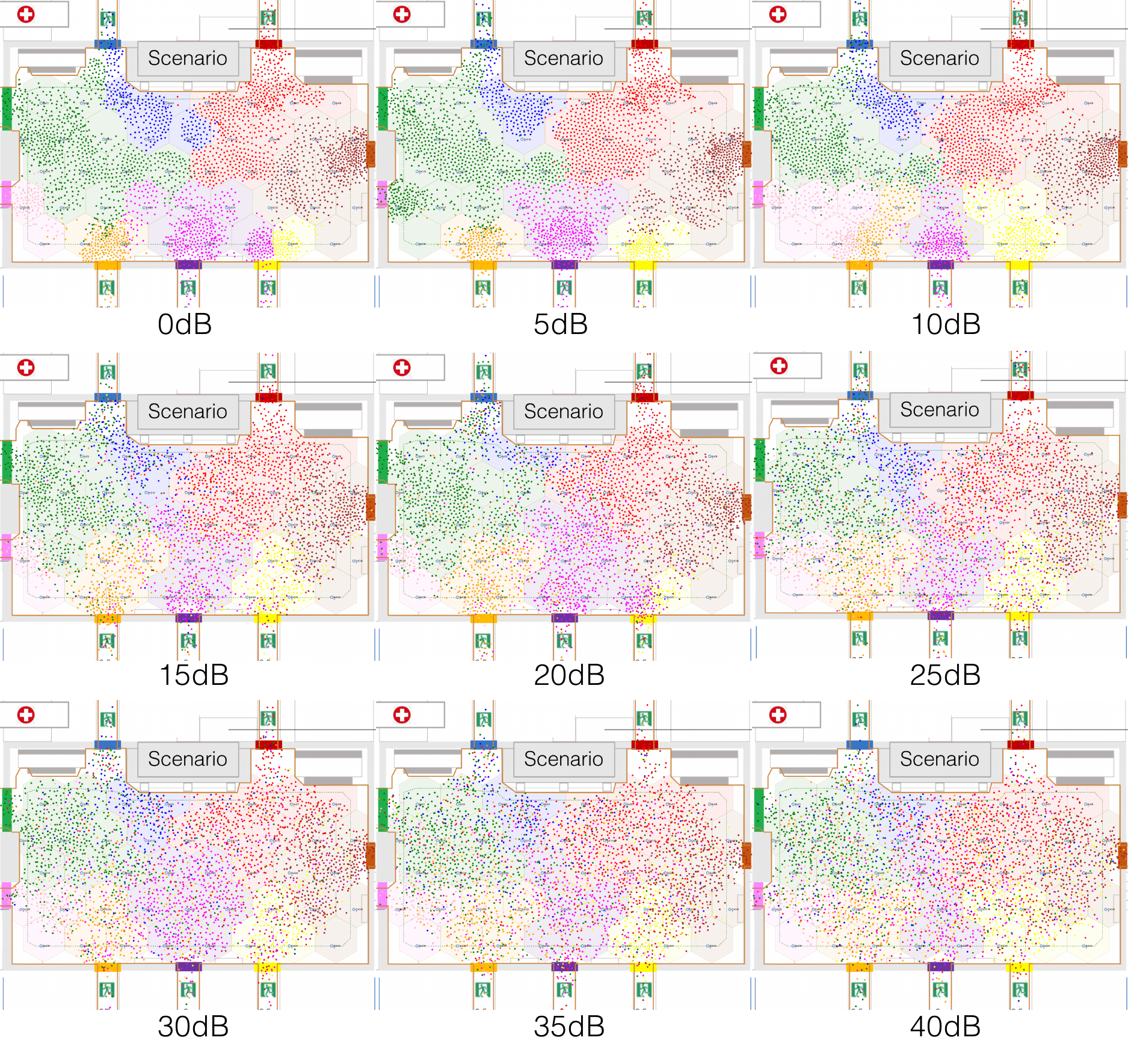}
\par\end{centering}
\caption{\label{fig:Snapshots}Still images $25$ seconds after the start of
the evacuation from single run simulation experiments for different
standard deviation values $\sigma_{g}$. The cells are shaded with
the exit-gate color allocated by the controller node. Colored dots
represent pedestrians with the colors shown by their LED wristbands. }
\end{figure}

For illustration purposes, Figure \ref{fig:Snapshots} shows still
images $25$ seconds after the start of the evacuation for different
standard deviation values $\sigma_{g}$, from $0$dB to $40$dB. As
expected, the snapshots exhibited a progressive level of error in
cell detection, becoming more evident from $15$dB. 

\begin{figure}
\begin{centering}
\includegraphics[width=0.6\columnwidth]{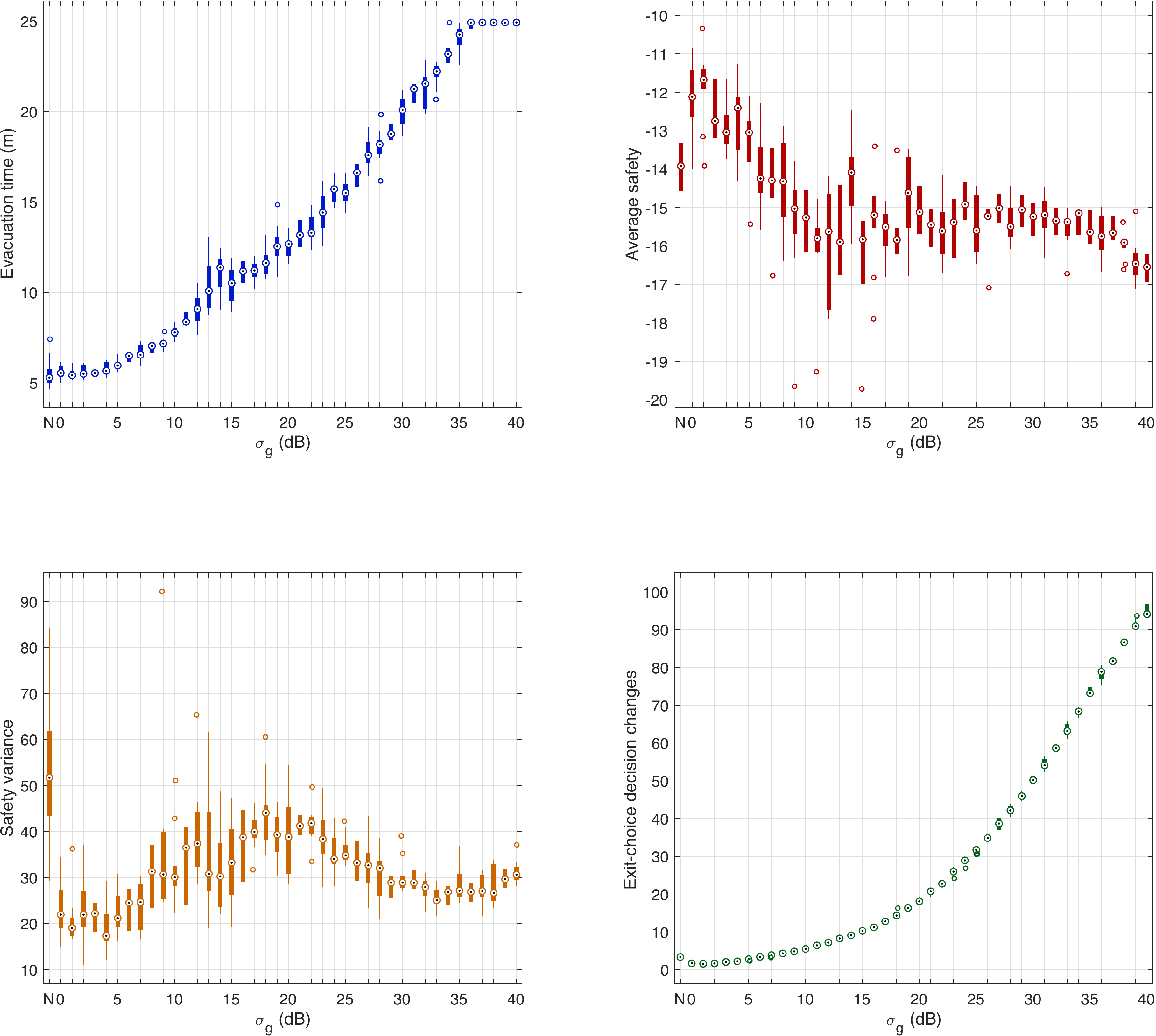}
\par\end{centering}
\caption{\label{fig:Sensitivity-analysis-of-NEF}Sensitivity analysis of the
positioning uncertainty in evacuation scenarios without external pedestrian
flows. The box-plots on the first row show the sensitivity of evacuation
time and average safety to standard deviation values $\sigma_{g}$
in the range $0$dB to $40$dB. The second-row plots show the sensitivity
of safety variance and the number of decision changes to the standard
deviation values $\sigma_{g}$. In the four box-plots, $\sigma_{g}=N$
represents an evacuation scenario in which pedestrians do not use
CellEVAC.}
\end{figure}

In evacuation scenarios without external flows (Figure \ref{fig:Sensitivity-analysis-of-NEF}),
results revealed that evacuation time increased linearly for $\sigma_{g}$
above $5$dB. Regarding safety, increasing values of $\sigma_{g}$
had a significant negative impact on average safety, though for $\sigma_{g}$
above $10$dB average safety stabilized around $-15$. Besides, the
impact on safety variance was not so significant as in average safety.
As expected, performance worsened for increasing $\sigma_{g}$, though
for values above $20$dB safety variance tended to improve and stabilize.
At the cost of an increasing evacuation time, we observed how uncoordinated
pedestrians' movement when positioning uncertainty was high, made
spatial-density at exit gates decrease, and so safety measurements
stabilize or improve. The number of exit-choice decision changes increased
exponentially with $\sigma_{g}$, due to the logarithmic scale (dB)
used to define the values of $\sigma_{g}$.

When compared to evacuations without external flows in which CellEVAC
did not operate, and not considering the safety variance, we observed
that using CellEVAC was beneficial strictly for values of $\sigma_{g}$
below $5$dB. Higher values of $\sigma_{g}$ could be valid at the
cost of an increase in evacuation time. However, note that not using
CellEVAC comes at the cost of a significantly higher safety variance.

\begin{figure}
\begin{centering}
\includegraphics[width=0.6\columnwidth]{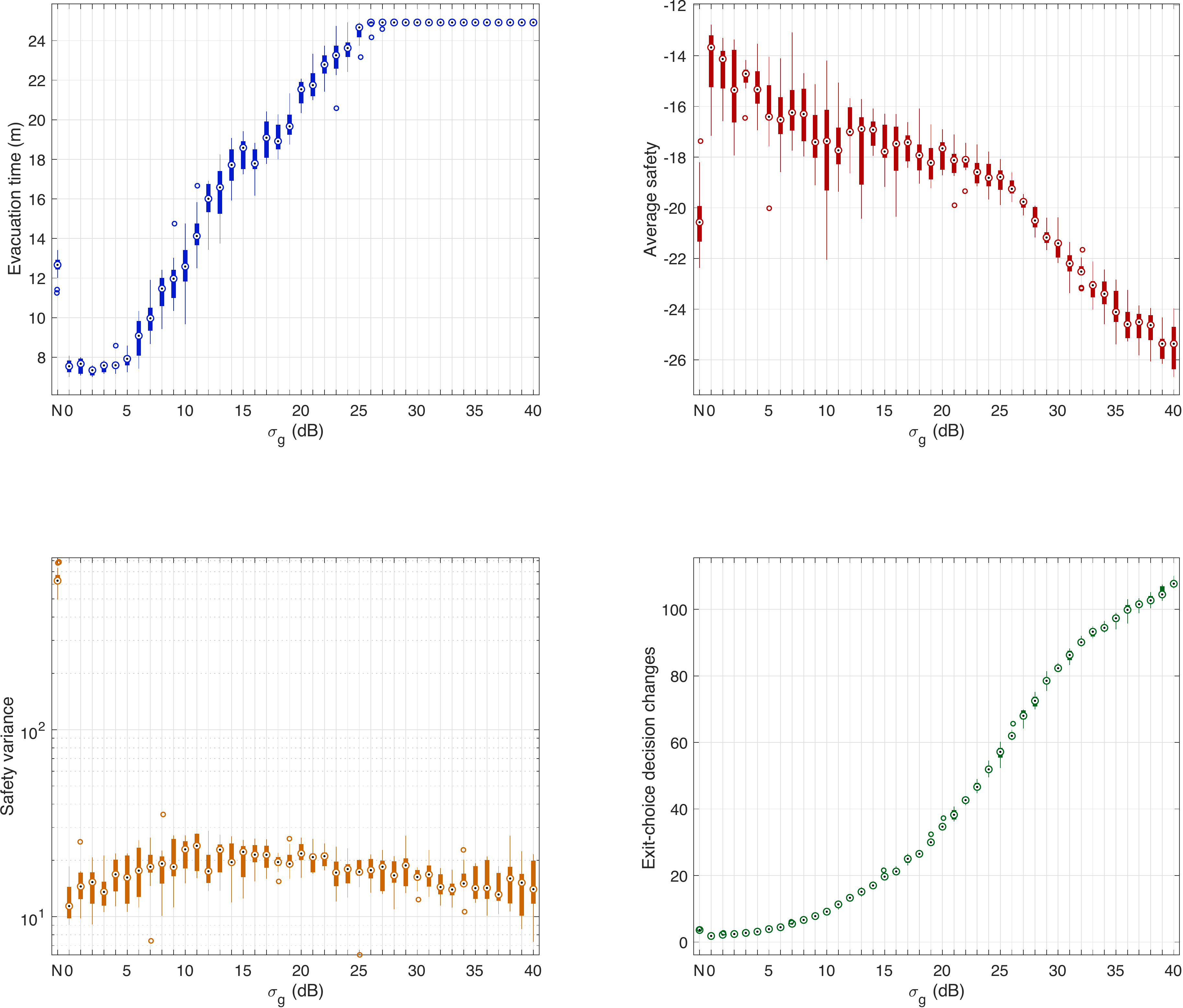}
\par\end{centering}

\caption{\label{fig:Sensitivity-analysis-of-EF}Sensitivity analysis of the
positioning uncertainty in evacuation scenarios with external pedestrian
flows. The box-plots on the first row show the sensitivity of evacuation
time and average safety to standard deviation values $\sigma_{g}$
in the range $0$dB to $40$dB. The second-row plots show the sensitivity
of safety variance and the number of decision changes to the standard
deviation values $\sigma_{g}$. In the four box-plots, $\sigma_{g}=N$
represents an evacuation scenario in which pedestrians do not use
CellEVAC.}
\end{figure}

In evacuation scenarios with external flows, the sensitivity analysis
results revealed the same trend as without external pedestrian flows
(Figure \ref{fig:Sensitivity-analysis-of-EF}). When compared to evacuations
that did not use CellEVAC, and not considering the safety variance,
the benefit of CellEVAC expanded to $\sigma_{g}$ below $10$dB. However,
note that safety variance is exceptionally high when not using CellEVAC,
which means that pedestrian flows at different exit gates is highly
unbalanced.

Overall, the results of the sensitivity analyses for scenarios with
and without external flows suggest a clear benefit of using CellEVAC
if the positioning system exhibits RSSI random variations below $10$dB.

\subsection{Optimizing CellEVAC for Different Positioning Uncertainty Levels}

\begin{figure}
\begin{centering}
\includegraphics[width=0.8\columnwidth]{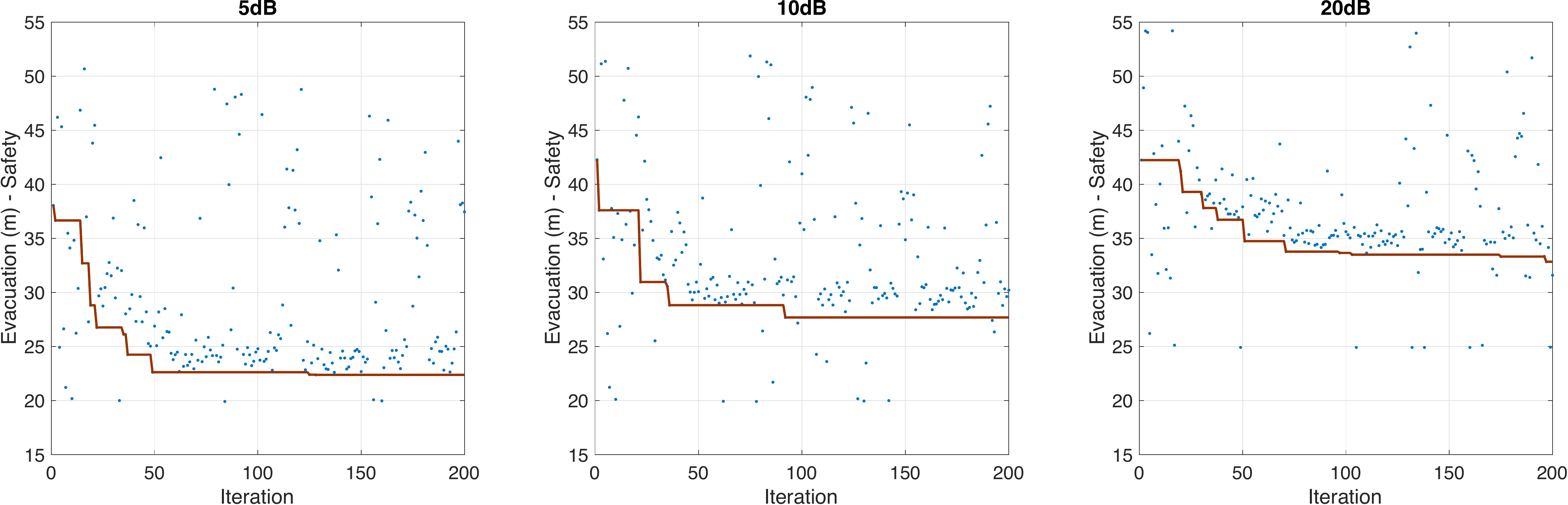}
\par\end{centering}
\caption{\label{fig:Optimization Cell}Progress of the Tabu-search simulation-optimization
of the MLM models that implement the CellEVAC guidance system for
$\sigma_{g}$ equal to $5$, 15 and 20dB. Solutions below the current
best solution (red line) correspond to non-viable solutions.}
\end{figure}

Figure \ref{fig:Optimization Cell} shows the progress of the Tabu-search
simulation-optimization of the MLM models' parameter configurations
for values of $\sigma_{g}$ equal to $5$, $15$ and $20$dB. The
objective was to optimize evacuation time and average safety in scenarios
with external flows. It was assumed that the entire population of
evacuees followed the indications of the CellEVAC system. Also, we
imposed an arbitrary simulation stop-limit of 25 minutes to evacuation
time, and a restriction to the viability of the solutions was incorporated
to remove solutions in which there were pedestrians pending evacuation.

The optimal parameter configurations found are shown in Table \ref{tab:Optimal-configuration-dBs}.
We did not observe significant differences between the different parameter
configurations, except for the $\beta_{D}$ and $\beta_{G}$ parameters.
The distance parameter gained more influence in scenarios with high
positioning uncertainty. Remarkably, for $\sigma_{g}=20$dB, the group
parameter $\beta_{G}$ had a positive sign. Our interpretation is
that a higher influence of distance and a positive value of $\beta_{G}$
contributes to more uniformity in the exit gate indications and less
scattering in color allocation to cells. As a consequence of this,
the probability of location error decreases.

\begin{table}
\caption{\label{tab:Optimal-configuration-dBs}Optimal parameter configurations
of the MLM model for different values of $\sigma_{g}$.}

\centering{}%
\begin{tabular}{cccccc}
\hline 
$\sigma_{g}$ & {\small{}$\beta_{D}$} & {\small{}$\beta_{G}$} & {\small{}$\beta_{E}$} & {\small{}$\beta_{W}$} & {\small{}$\beta_{C}$}\tabularnewline
\hline 
0dB & {\small{}-17.723} & {\small{}-2.181} & {\small{}-1.671} & {\small{}1.064} & {\small{}2.594}\tabularnewline
{\small{}5dB} & -16,040 & -3,224 & -2,267 & 0 & 6,816\tabularnewline
{\small{}10dB} & -17.696 & -2 & -2 & 1.685 & 3\tabularnewline
20dB & -28.479 & 10 & -3.083 & 0.041 & 4.025\tabularnewline
\hline 
\end{tabular}
\end{table}

\begin{figure}
\begin{centering}
\includegraphics[width=1\linewidth]{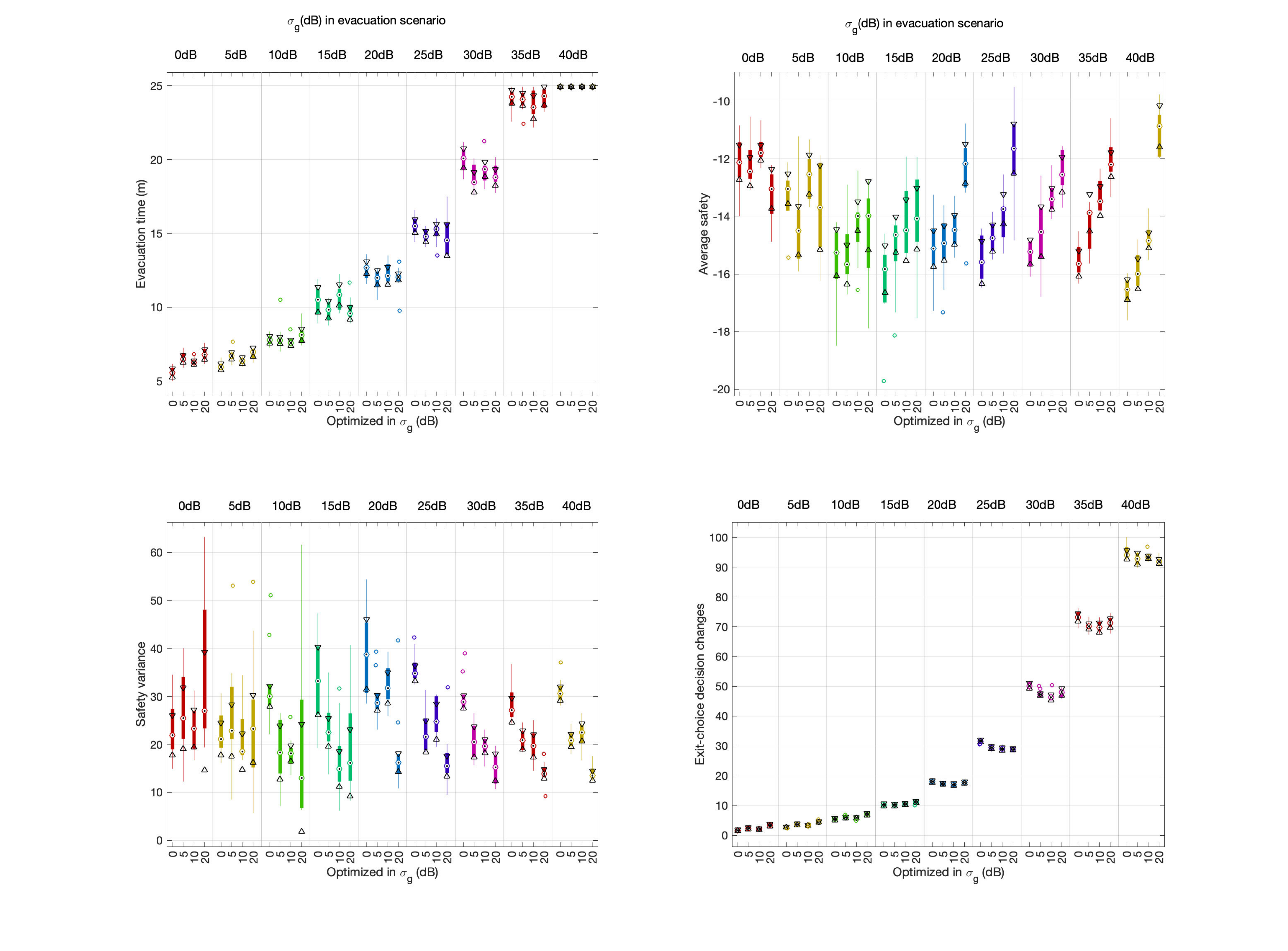}
\par\end{centering}
\caption{\label{fig:Boxplots-NEF-Performance}Median box-plots of the performance
measurements of the optimal configurations of CellEVAC for $\sigma_{g}=0,5,10,20$dB,
in evacuation scenarios without external pedestrian flows and with
positioning uncertainty levels from $0$dB to $40$dB in steps of
$5$dB. Bottom horizontal axes categorize the optimal configuration
of parameters used ($0$, $5$, $10$, or $20$dB) to configure CellEVAC.
Upper horizontal axes categorize the $\sigma_{g}$ value that models
the positioning system in the evacuation scenario. For instance, a
value of $25$dB in the axis ``$\sigma_{g}$(dB) in evacuation scenario''
and $15$dB in ``Optimized in $\sigma_{g}$(dB)'' expresses that
CellEVAC has been configured to use the optimal configuration found
with $15$dB, and that it is tested in an evacuation scenario with
$\sigma_{g}=25$dB. The triangles in the box-plots display the variability
of the median between samples. If the notches of two measurements
do not overlap, they have different medians at the 5\% significance
level. }
\end{figure}

\begin{figure}
\begin{centering}
\includegraphics[width=1\columnwidth]{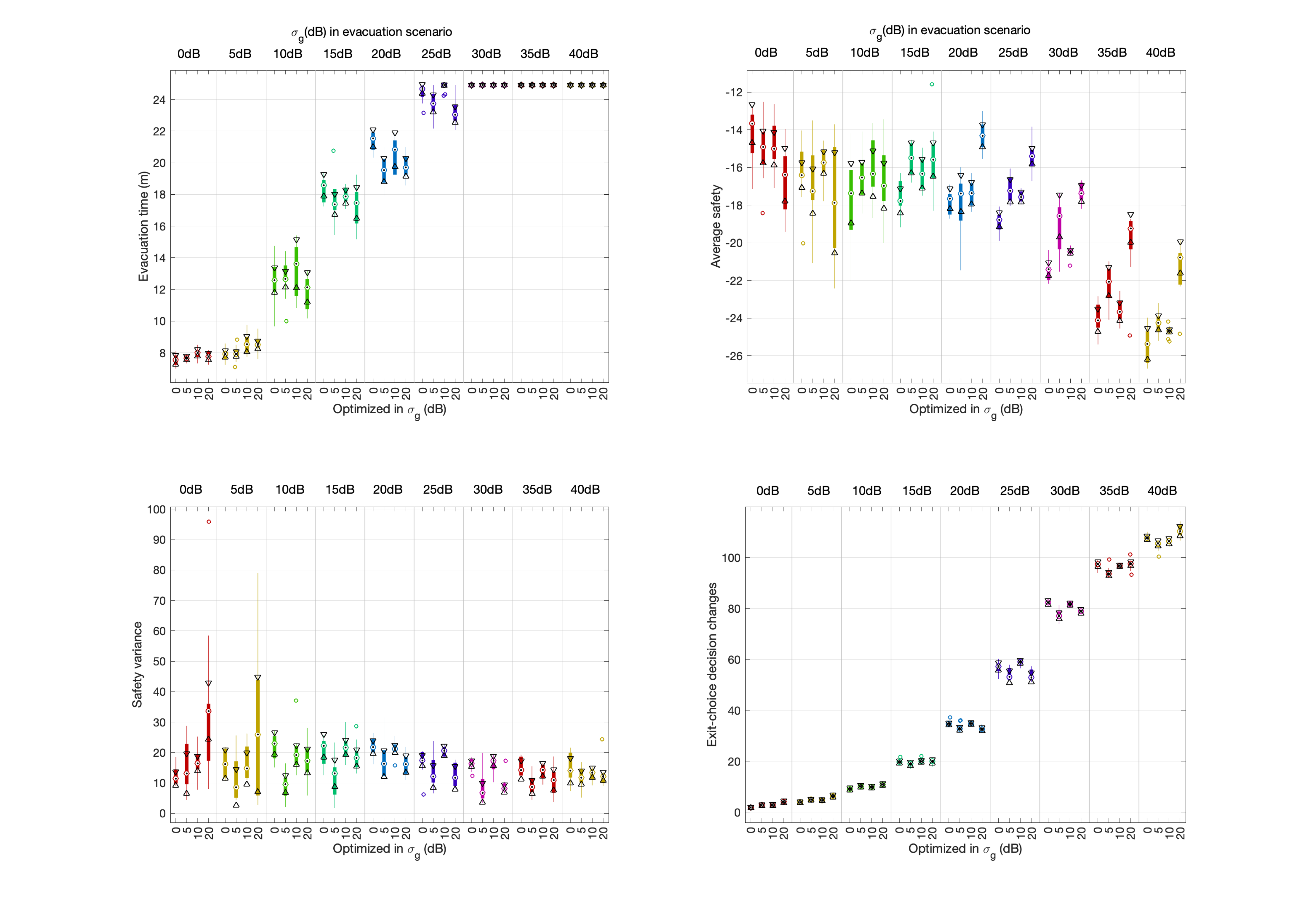}
\par\end{centering}
\caption{\label{fig:Boxplots-EF-Performance}Median box-plots of the performance
measurements of the optimal configurations of CellEVAC for $\sigma_{g}=0,5,10,20$dB,
in evacuation scenarios with external pedestrian flows and with positioning
uncertainty levels from $0$dB to $40$dB in steps of $5$dB. Bottom
horizontal axes categorize the optimal configuration of parameters
used ($0$, $5$, $10$, or $20$dB) to configure CellEVAC. Upper
horizontal axes categorize the $\sigma_{g}$ value that models the
positioning system in the evacuation scenario.}
\end{figure}

The optimal configurations found were tested in evacuations with different
positioning uncertainty levels, from $0$dB to $40$dB in steps of
$5$dB. The results have been summarized in Figures \ref{fig:Boxplots-NEF-Performance}
and \ref{fig:Boxplots-EF-Performance} under evacuation scenarios
without and with external flows, respectively.

In evacuation scenarios without external flows (Figure \ref{fig:Boxplots-NEF-Performance}),
average evacuation time, and exit-choice decision change performance
measurements did not show significant differences between the different
configurations and evacuation scenarios. Interestingly, we found a
positive trend in the results in terms of average safety and safety
variance for evacuation scenarios for $20$dB and above when using
the optimal configuration found for $20$dB. In evacuation scenarios
with external flows (Figure \ref{fig:Boxplots-EF-Performance}), the
results were similar except for safety variance, for which we did
not observe any improvement.

The results presented in section \ref{subsec:Sensitivity-Analysis-of}
show that CellEVAC is useful only if the positioning system exhibits
RSSI random variations below $10$dB. Besides, the performance analysis
results of the optimal configurations do not exhibit any improvement
below $20$dB. Consequently, we can conclude that there is no evidence
that optimizing the MLM model under the assumption of an expected
random variance of RSSI contributes to an improvement in the performance
of the CellEVAC system.

\section{\label{sec:Discussion}Conclusion}

Our use of an MLM model to implement the control logic of Cell-based
crowd evacuation systems has proven to be very efficient (see \cite{lopez-carmonaCellEVACAdaptiveGuidance2020}).
However, as in other existing works on cell-based crowd evacuation
systems \cite{Abdelghany20141105,Zhong20141125,wongOptimizedEvacuationRoute2017},
little attention has been paid to propose a system architecture based
on existing technologies and assuming real communication and positioning
infrastructures. In our opinion, these considerations are crucial
to boost the real implementation of these systems. 

In this paper, we have proposed a specific system architecture built
upon radio-controlled LED wristbands that connect with a cell-node
network and a controller node, through radio-frequency communication
channels. The use of LED wristbands has numerous advantages, among
which we highlight its low cost, being a technology widely used to
create visual effects at large events, and being an intuitive and
straightforward interface that can make it exceptionally efficient
in emergencies. This type of indication system greatly simplifies
the interpretation of exit gate indications, which is particularly
important in stressful situations found typically in evacuation scenarios.
Indirectly, it can also improve the accuracy of the detection of pedestrian
flows in the controller node.

Another of our aims was to quantitatively study the sensitivity of
evacuation time and safety to uncertainty in the positioning system.
With this objective, we have modeled the communication channel between
the LED wristbands and the cell-nodes using a log-normal propagation
model. To model different levels of uncertainty in positioning, we
have modulated the random variations of RSSI from the propagation
model. In the sensitivity analysis of performance parameters to different
values of RSSI variance, CellEVAC is shown to be operational strictly
up to values of 10dB. The system generates too many color changes
in the wristbands and a significant increase in evacuation times for
higher values.

Our last goal was to evaluate if it was possible to improve the CellEVAC
performance obtaining new optimal MLM parameter configurations in
which different levels of RSSI standard deviation were assumed. The
results obtained have revealed that improvements found are relevant
only for evacuation scenarios with levels of positioning uncertainty
greater than 20dB, in which CellEVAC is not operational. Thus, to
optimize the MLM model used in the CellEVAC control logic, it is valid
to assume that the positioning system is error-free. However, the
system cannot be applied in a real environment if the standard deviation
of the RSSI values is greater than 10dB.

Several extensions are being considered for this research, mainly
focused on investigating how to expand the allowed range of RSSI variation
without the need to modify the existing positioning infrastructure.
Another research extension is related to developing a prototype of
the CellEVAC system architecture proposed in this paper.

\authorcontributions{Conceptualization, M.A.L.C; methodology, M.A.L.C and A.P.; software,
M.A.L.C.; validation, A.P. and M.A.L.C; investigation, M.A.L.C and
A.P.; resources, M.A.L.C; data curation, M.A.L.C and A.P.; writing--original
draft preparation, M.A.L.C.; writing--review and editing, M.A.L.C
and A.P.; visualization, A.P. and M.A.L.C; supervision, M.A.L.C; project
administration, M.A.L.C; funding acquisition, M.A.L.C.}

\funding{This work was supported in part by the Spanish Ministry of Economy,
Industry, and Competitiveness under Grant TIN2016-80622-P (AEI/FEDER,
UE). }

\conflictsofinterest{The authors declare no conflict of interest.}





\end{document}